# Magnetoresistance in Thin Permalloy Film (10nm-thick and 30-200nm-wide) Nanocontacts Fabricated by e-Beam Lithography.

Nicolás. García[1], Cheng Hao[1], Lu Yonhua[1], Manuel. Muñoz[1], Yifang Chen[2], Zhengqi Lu[3], Yun Zhou, Genhua Pan[3], Zheng Cui[2] and A. A. Pasa[4]

1 Laboratorio de Física de Sistemas Pequeños y Nanotecnología, Consejo Superior de Investigaciones Científicas (CSIC), Serrano 144, 28006 Madrid (Spain)
2 Rutherford Appleton Laboratory, Chilton, Didcot, Oxon, UK OX11 0QX
3 School of Computing, Communication and Electronics, University of Plymouth, Plymouth, Drake Circus, Plymouth, Devon, UK PL4 8AA
4 Laboratório de Filmes Finos e Superfícies, Departamento de Física, Universidade Federal de Santa Catarina, Caixa Postal 476, 88.040-900 Florianópolis, SC, Brazil

ABSTRACT

In this paper we show spin dependent transport experiments in nanoconstrictions ranging from 30 to 200nm. These nanoconstrictions were fabricated combining electron beam lithography and thin film deposition techniques. Two types of geometries have been fabricated and investigated. We compare the experimental results with the theoretical estimation of the electrical resistance. Finally we show that the magnetoresistance for the different geometries does not scale with the resistance of the structure and obtain drops in voltage of 20mV at 20Oe.

Nowadays technologies for communications, automotive, medical and other areas of interest are demanding more and more nanometer size integrated structures. One of the fields that lately have taken more attention is spintronics.. The discovery of giant magnetoresistance using magnetic and non-magnetic multilayers(1,2) represented a change. These systems work with a good stability because the materials are all metallic and conduction is good so that reasonable currents (mA) can be kept stable. Also giant magnetoresistance appears in granular materials (3,4). Real and useful devices in the micrometer size have been developed by industry. Moreover, it is general accepted that it will be ideal to have devices in the Terabit/$cm^2$, that is an ultimate integration. For that science and technology needs spintronics units of nanometer size and large changes of voltage drop when the magnetic field is switched. Regarding this point there is a proposal by Garcia and collaborators based in ballistic magnetoresistance (BMR) in atomic contacts (5-7) and nanocontacts (8-10). There have been challenging that the large MR observed in nanocontact is due to contact modification. Because of this it is interesting to perform experiments in thin film nanoconstrictions.

In this letter we present experimental study of the MR of magnetic nanoconstriction fabricated by e-beam lithography and thin film deposition to clarify some previous issues. This study demonstrates that the MR effect is not and cumulative effect and that the MR is a superposition of positive and negative values added up in the through the length of the device. We find values up to 2% in single Py films 10nm thick. However we have changes of resistance of as much as 8Ω with up to 2mA DC current that implies drop voltages of 16mV with 20 Oe field and up to 28mV when we apply 2 mA AC current at 200 Hz. Also the MR in percentage is low because the leads have a large resistance but most of the MR comes up from the contact region.

The devices studied in this work were fabricated by electron beam lithography and thin film deposition. For the lithography step, the substrates were covered with double layer of LOR/PMMA resists using spin coating and then baked in an oven. The bottom layer: 30% LOR 3A was spun at 2000rpm for 60 seconds depending on desired resist height. Subsequently, the resist was baked for 20minutes in an oven at 180ºC. Top layer: 2.5% PMMA 2041 was spun at 1800 rpm for 60 seconds. For the highest possible resolution, hardness and best resist profile, a time of 1 hour was used for baking in an oven at 180ºC. It is found that the thickness of the bilayer is critical for lift–off. The typical thickness is that LOR is about 60nm and the PMMA is 120nm. The sample was exposed using e-beam writer VB6 from Leica Cambridge Ltd at 100 keV with dose range from 500 -2000 $\mu C/cm^2$. Subsequently a wet development step was carried out for 60 seconds in a 3:1 mixture of IPA:MIBK and flushed in IPA, then put in a 3:2 mixture of CD26:DI water for 60 seconds, finally flushed in DI water.

After forming the mask, Ti (2nm)/NiFe (5~15nm)/Ti (2nm) thin films were deposited in a Nordiko sputtering system. For the lift-off, the sample was immersed in 1165 resist remover at 86ºC until all residual material has been lifted off. Then it is flushed in acetone, and then in IPA. The sample was dried by $N_2$ gas. Then the wafer was coated with LOR and 1805 photoresist bilayers after nano-constriction pattern was made. Bottom layer LOR 3B was spun at 4000 rpm for 60 secondµ30 minutes in an oven at 95ºC. Finally, the sample was exposed using a maskaligner. Subsequently a wet development step was carried out for 12 seconds in a 1:5 mixture of 2401:DI water and flushed in DI water.

We have fabricated two kinds of samples whose structure are studied by SEM and show in figure 1. To fabricate the nanoconstriction we use two triangles facing each other and the size of the constriction is controlled by the employed dose in the lithography step. With this technique we have fabricated contacts from 30 to 250 nm in width. The main difference between the two types of samples is the distance between the Au pads which allow electrical contact to the sample. While in the type 1 (figure 1a) we pattern the electrical pads as closer as possible to the nanoconstriction, in type 2 (figure 1b) we use two microfabricated lines of the same material as the contact to conned the pads to the nanoconstriction. These lines are 30 µm in length and 9 µm in width. The thickness of the Py layer is 10 nm according to the calibration of the sputtering system. A commercial Digital Instruments scanning probe microscope was used to perform AFM measurements to study the configuration of the contacts.

In order to estimate the resistivity of the Py film, we have grown continuous film of 10 nm thickness and 0.2x0.8 cm as lateral dimensions. The measured resistance for such films is 39 Ohms which lead us to a film resistivity $\rho=1.56.10^{-4}$ $\Omega$.cm. This value of the resistivity is one order of magnitude than that for bulk Py.

The ohmic resistance for each type of contact can easily be calculated. If we take the geometry shown in Fig 2 we have that for the contacts type 1 the resistance is given by:

$$R = \frac{\rho}{2t \tan \alpha} \left[ \ln\left( \frac{(l-a)\tan \alpha + w/2}{w/2} \right) + \ln\left( \frac{a \tan \alpha + w/2}{w/2} \right) \right] \quad [1]$$

while for the contacts type 2 the expression for R is:

$$R = \frac{\rho}{t}\left[\frac{L}{W} + \frac{1}{\tan\alpha}\ln\left(\frac{l\tan\alpha + w/2}{w/2}\right)\right] \quad\ldots\ldots\ldots\ldots\ldots\ldots[2]$$

where t is the thickness of the Py film; L and W are respectively the length and width of the Py bars that connect the leads to the constriction. The logarithm terms comes up from the triangular shapes facing each other that conform the constriction at the end.

Typical curves of the resistance as a function of the applied field (H) are shown in Fig 3. The resistance is measured while applying a constant current of 0.2 to 2mA and the magnetic field is applied in the plane of the film and perpendicular to the current. Figure 3 (a) shows the MR for the continuous Py film. In Fig 3 (b) we can see the MR curve for one contact of type 1.These measurements were performed at RT (top of figure 3 (c)) and at 80 K (bottom) so that we can discard any possible leak of current through the Si substrate. Fig. 3 (c) shows the MR of one contact of type 2 at room temperature.

As expected, the measured resistance of the contact type 1 is smaller than for the contact type 2, where as well as the contribution to the resistance from the constriction, we have the contribution from the long Py bars. However, the change in resistance (MR) is bigger for the contacts type 1 than for contacts type 2. What we would expect to have for these contacts is that the MR is given by the change in the resistance coming from the constriction (similar to type 1 contacts) plus the change in resistance due to the long bars. The fact that less resistive contacts have bigger values of the MR than more resistive contacts, implies that the magnetoresistance is not a cumulative effect. Or let us put in another way; this indicates that the large change in MR comes from the contact region, however the percentage of MR is masked by the large resistance of the cones leads connecting the nanocontact. In addition to this for the longer samples (type 2) we have a subtracting part to the MR rising in the nanocontact produced by the bars that connect the cones. We have not determined yet the MR of the contact and of course the structure that has been fabricated is nonballistic obviously. We need to have the pads very near to the nanocontact, a few nanometer and not 1000´s of nm as we have now. This is a difficult task and we have to devise other alternatives to clear this point. But one thing we have clear is that if we were able to reduce the resistance of the leads the MR percentage will be very large. The percentage of MR produced at the constriction is masked and reduces by the large resistance of the leads connecting the nanocontacts and the pads. If we estimate the resistance of the contact (see right ordinates axis of Fig.3b) and assume that the change of resistance comes from the contact the percentage of MR will be very large. The devices depicted here are not ballistic but could be made by allocating the pads at distances smaller than 20 nm from the constriction. Work is oriented in this direction. For the moment we cannot say what is the contribution of the constriction region to the MR. In any case work on ballistic transport in metallic nanostructures ,may have to be reconsidered in view that many earlier references of importance where overseen (12).

In figure 4 we present the experimental results for the set of samples type 1. Left ordinates represent the resistance R (■) versus the contact width $W_c$. In the right ordinates we show the magnetoresistance $\Delta R$ versus the constriction width (▲). We remark how the resistance does not depend on the contact width. This denotes that the contact does not determine the total resistance, and that the resistance of the contact must be much smaller. The continuous line in figure 4 is the resistance in function of the contact width given by expression 1 where we use using l=1000 nm; a=l/2 α=30º and t=10 nm as geometrical parameters for the contact.

As mentioned before the samples where studied by AFM and the dimensions obtained with these measurements are introduced in Formula 1. These results are shown in Figure 4 (○).

It should be stressed that for the technology what is measured is the voltage drop (voltage signal) under application of magnetic field. These samples can admit 2mA current in DC and 4 mA in 5ms pulse AC, that provide voltage drops 14mV and 28mV for a 20Oe coercive field. These voltages drops and currents are possible, following our observations, because the growth is done directly on Si by using the Schottky barrier to prevent the shunting of the current. This is an important thing because if the samples are grown in glass we can only apply currents of 0.1mA. This is due to the fact that thermal conductivity is 100 times larger in Si than in glass and the dissipation of heat is an important issue for current stability. The results are also stable up to 400K and the device is nanoscopic.

In conclusion, combining e-beam lithography and thin film deposition, two types of nanometer size constrictions have been fabricated. We have performed experiments on spin dependent transport of such contacts showing that the resistance is independent of the width of the constriction. Furthermore, we showed that the change in resistance is bigger for the less resistive samples than for the more resistive ones. These results suggest than the MR is not a cumulative effect and is rising from the region contact. Furthermore, there is a proof that the MR of a film is much smaller than the MR of constriction (compare Fig.3a with 3b. In order to discriminate the value of the BMR we need to locate the pads for measurement at nanometer distance from the nanocontact, then the BMR in percentage may be very large. More work is being done in this line.

This work has been supported by the BMR-NMP-EU project. We thank to our partner Lenar Tagirov for discussions.

**Figure Captions.**
-------------------------------------------------------------------------------------------------

**Figure 1.** SEM micrographs for the nanofabricated constrictions: a) Constriction type 1 where the pads are patterned as close as possible to the constriction, and b) contacts type 2 where the two Py bars connect the constriction to the pads.

**Figure 2**. Geometry of contacts type 1 used to calculate the ohmic resistance.

**Figure 3.** a) magnetoresistance of a continuos Py film 10 nm thick, notice the small values of $\Delta R$ and MR. b) MR for the constrictions type 1. The changes in $\Delta R$ are 600 larger and the same is the estimated MR. We notice that $\Delta R$ does not depend much on T neither of the bias voltage. Contrarily for the films without constriction the $\Delta R$ is three times larger for the parallel field than for the perpendicular field. c) MR for a constriction type 2. Note how been R much bigger than for type 1 contacts, $\Delta R$ is smaller.

**Figure 4.** Plot of the resistance versus the contact width $W_c$. is represented in left ordinates. (■) are the experimental values and (○) calculated with formula 1 using the dimensions of the contacts obtained with AFM. Right ordinates represent the measured values of $\Delta R$ (▲). Thick line is the ohmic resistance versus contact width given by formula 1.

Figure 1

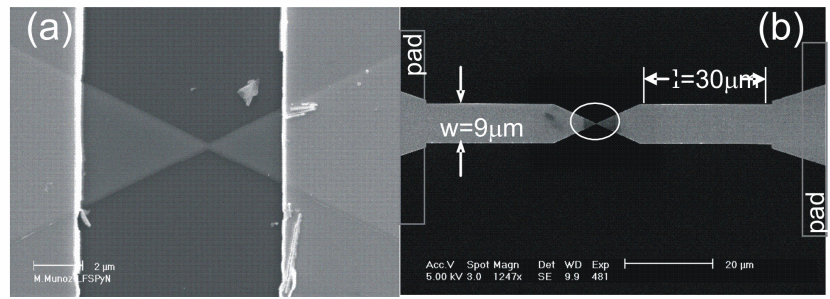

Figure 2

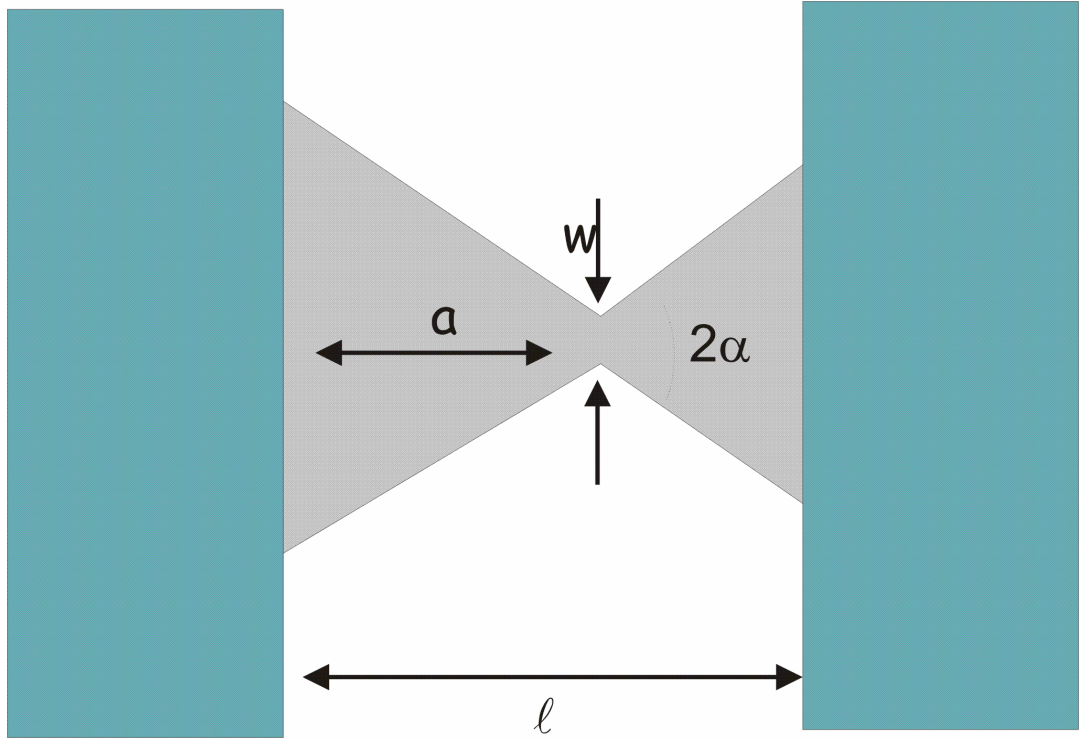

Figure 3

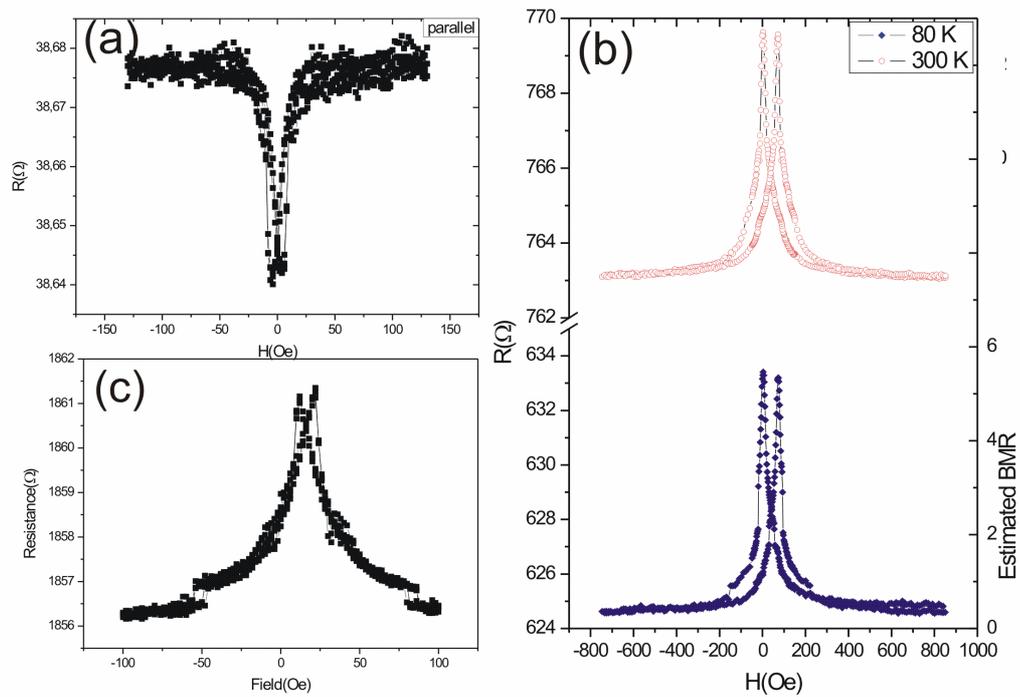

Figure 4

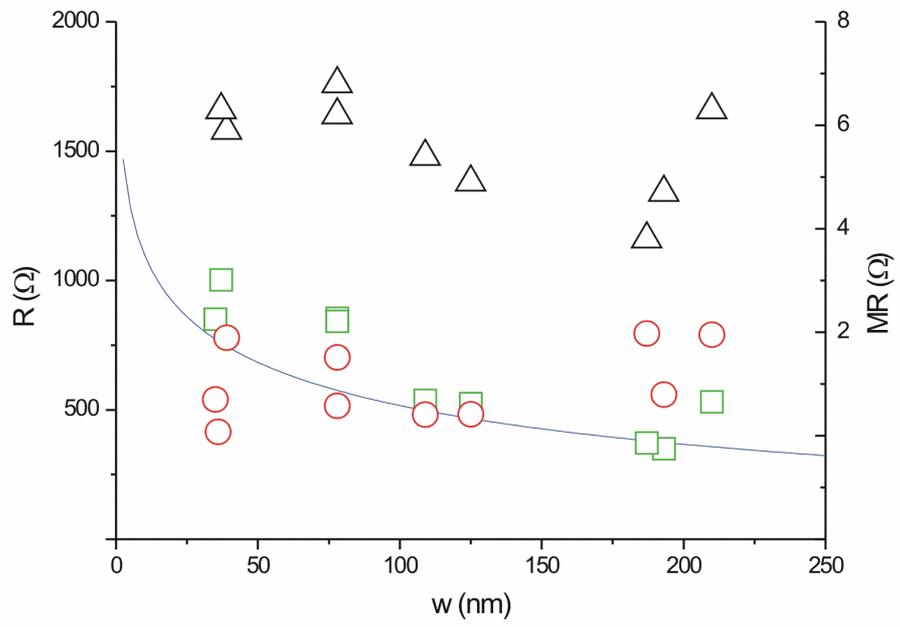